\documentclass[conference]{IEEEtran}
\ifCLASSINFOpdf
\else
   \usepackage[dvips]{graphicx}
\fi
\begin{document}
%
\title{1-State Error-Trellis Decoding of LDPC Convolutional Codes Based on Circulant Matrices}

\author{\IEEEauthorblockN{Masato Tajima}
\IEEEauthorblockA{Graduate School of Sci. and Eng.\\
University of Toyama\\
3190 Gofuku, Toyama 930-8555, Japan\\
Email: tajima@eng.u-toyama.ac.jp}
\and
\IEEEauthorblockN{Koji Okino}
\IEEEauthorblockA{Information Technology Center\\
University of Toyama\\
3190 Gofuku, Toyama 930-8555, Japan\\
Email: okino@itc.u-toyama.ac.jp}
\and
\IEEEauthorblockN{Takashi Miyagoshi}
\IEEEauthorblockA{Graduate School of Sci. and Eng.\\
University of Toyama\\
3190 Gofuku, Toyama 930-8555, Japan\\
Email: miyagosi@eng.u-toyama.ac.jp}}


%


\maketitle

\begin{abstract}
We consider the decoding of convolutional codes using an error trellis constructed based on a submatrix of a given check matrix. In the proposed method, the syndrome-subsequence computed using the remaining submatrix is utilized as auxiliary information for decoding. Then the ML error path is correctly decoded using the degenerate error trellis. We also show that the decoding complexity of the proposed method is basically identical with that of the conventional one based on the original error trellis. Next, we apply the method to check matrices with monomial entries proposed by Tanner et al. By choosing any row of the check matrix as the submatrix for error-trellis construction, a $1$-state error trellis is obtained. Noting the fact that a likelihood-concentration on the all-zero state and the states with many $0$'s occurs in the error trellis, we present a simplified decoding method based on a $1$-state error trellis, from which decoding-complexity reduction is realized.
\end{abstract}


%
\IEEEpeerreviewmaketitle

\section{Introduction}
Tanner et al. [10] presented a class of algebraically constructed quasi-cyclic (QC) LDPC codes and their convolutional counterparts. Owing to their construction, check matrices of obtained LDPC convolutional codes have monomial entries and then each column (row) has a common factor of the form $D^l$. On the other hand, Ariel and Snyders [1] showed that when some ``column'' of a polynomial check matrix $H(D)$ has a factor $D^l$, there is a possibility that state-space reduction can be realized. For the same case (i.e., some column of $H(D)$ has a factor $D^l$), the authors [9] showed that the results of [1] can be equally obtained using shifted error-subsequences. These ideas can be directly applied to check matrices with monomial entries obtained from the construction of Tanner et al. Actually, $H(D)$ can be modified as $H'(D)$ with the error-correcting capability being preserved. Let $H''(D)$ be the factored-out version of $H'(D)$. Then we [8] showed that the state-space complexity of the error trellis based on $H''(D)$ can be controlled to some extent. However, the overall constraint length of $H''(D)$ is still large and therefore the use of ``trellis-based'' decoding is not feasible. Hence, another complexity reduction method is required for trellis-based decoding. In this paper, we present a decoding method using an error trellis constructed based on a ``submatrix'' of a given check matrix. Note that since some of check conditions are not taken into account, the degenerate error trellis contains additional error paths not allowed in the original error trellis. We show that the ML error path can be correctly decoded using the degenerate error trellis, if the syndrome-subsequence computed from the remaining submatrix is utilized as side information for decoding. In particular, consider check matrices proposed by Tanner et al. [10]. If we take any row for the purpose of error-trellis construction for decoding, each column (i.e., entry) of the row has a factor of the form $D^l$. Then factoring out these factors, a particular submatrix (i.e., row) with all $1$ entries is obtained and the number of states of the corresponding error trellis is one. We propose a sub-optimal decoding algorithm based on a 1-state error trellis. Applying the proposed method, a considerable decoding-complexity reduction is realized compared to the conventional one.

\section{Error Trellis Constructed Based on a Submatrix of a Check Matrix}
Consider an $(n, n-m)$ convolutional code $C$ over $GF(2)$ defined by a canonical [5], [6] check matrix $H(D)$ of size $m\times n$. Let $\nu$ be the overall constraint length of $H(D)$. Denote by $H^T(D)$ ($T$ means transpose) the corresponding syndrome former. Assume that $H^T(D)$ has the form
\begin{equation}
H^T(D)=(H_1^T(D), H_2^T(D)) ,
\end{equation}
where the size of $H_i(D)~(i=1, 2)$ is $m_i\times n$ ($m=m_1+m_2$) and the overall constraint length of $H_i(D)~(i=1, 2)$ is $\nu_i$ ($\nu=\nu_1+\nu_2$). Assume that $H_i(D)~(i=1, 2)$ is also canonical. In this case, for the time-$k$ error $\mbox{\boldmath $e$}_k$ and $\mbox{\boldmath $\zeta$}_k=(\mbox{\boldmath $\zeta$}_k^{(1)}, \mbox{\boldmath $\zeta$}_k^{(2)})$, we have
\begin{eqnarray}
\mbox{\boldmath $e$}_kH^T(D) &=& (\mbox{\boldmath $e$}_kH_1^T(D), \mbox{\boldmath $e$}_kH_2^T(D)) \nonumber \\
&=& (\mbox{\boldmath $\zeta$}_k^{(1)}, \mbox{\boldmath $\zeta$}_k^{(2)}) .
\end{eqnarray}
Here, it is assumed that given a received data $\mbox{\boldmath $z$}=\{\mbox{\boldmath $z$}_k\}$, the equation $\mbox{\boldmath $z$}_kH_i^T(D)=\mbox{\boldmath $\zeta$}_k^{(i)}~(i=1, 2)$ holds. The above relation implies that the original error path $\{\mbox{\boldmath $e$}_k\}$ associated with $H^T(D)$ can be represented using the error trellis based on either $H_1^T(D)$ or $H_2^T(D)$.

\begin{figure}[tb]
\begin{center}
\includegraphics[width=6.0cm,clip]{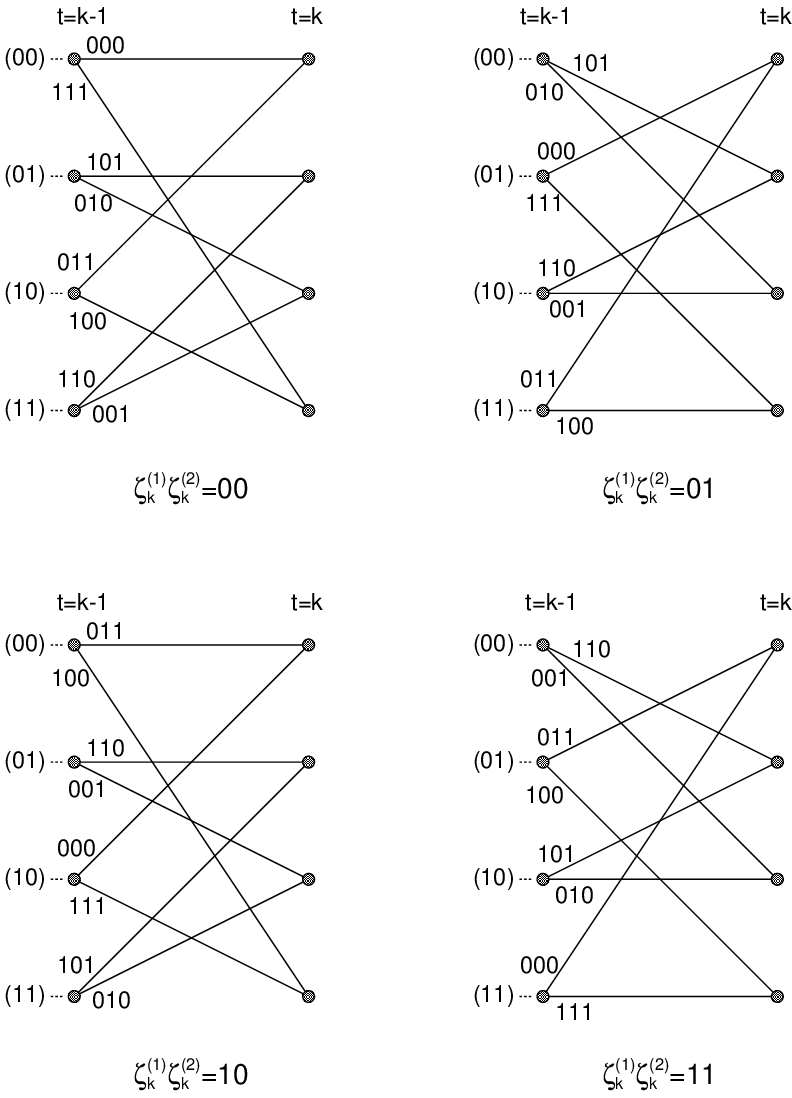}
\end{center}
\caption{The error-trellis modules associated with $H^T(D)$.}
\label{Fig.1}
\end{figure}
\begin{figure}[tb]
\begin{center}
\includegraphics[width=8.0cm,clip]{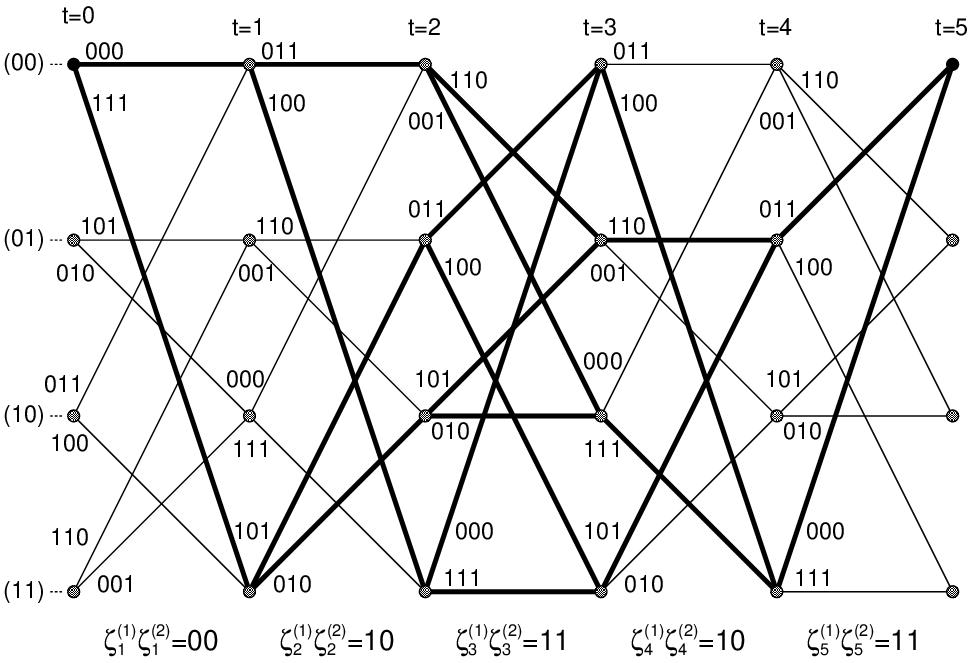}
\end{center}
\caption{An example error trellis based on $H^T(D)$.}
\label{Fig.2}
\end{figure}

For example, consider the $(3, 1)$ convolutional code $C$ defined by the check matrix 
\begin{equation}
H(D)=\left(
\begin{array}{ccc}
1+D& D& 1+D \\
D& 1& 1 
\end{array}
\right) .
\end{equation}
An error trellis of $C$ is constructed by concatenating the error-trellis modules [1] associated with $H^T(D)$. The set of four error-trellis modules associated with $H^T(D)$ and an example error trellis are depicted in Fig.1 and Fig.2, respectively. It is assumed that the corresponding code trellis is terminated in the all-zero state at time $5$. Hence, the error trellis in Fig.2 is terminated in state $(00)$, which corresponds to the syndrome-former state $\mbox{\boldmath $\sigma$}_5=(\sigma_5^{(1)}, \sigma_5^{(2)})$ (cf. Fig.5). From Fig.2, we have eight admissible error paths. $\mbox{\boldmath $e$}=000~100~000~100~000$ corresponds to the ML error path {\boldmath $\hat e$}.

\begin{figure}[tb]
\begin{center}
\includegraphics[width=6.0cm,clip]{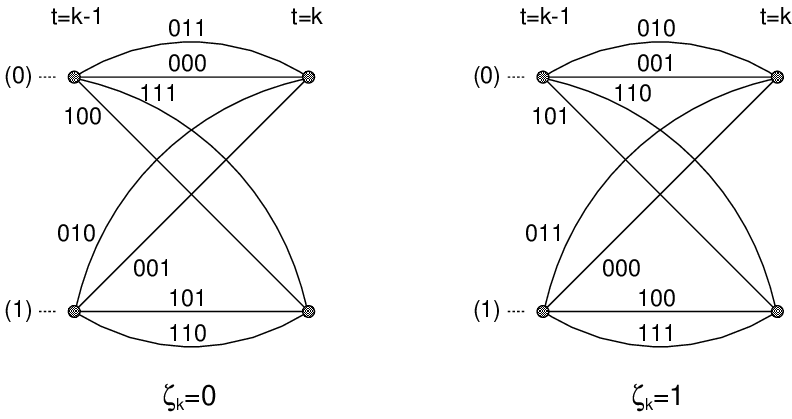}
\end{center}
\caption{The error-trellis modules associated with $H_2^T(D)$.}
\label{Fig.3}
\end{figure}
\begin{figure}[tb]
\begin{center}
\includegraphics[width=8.0cm,clip]{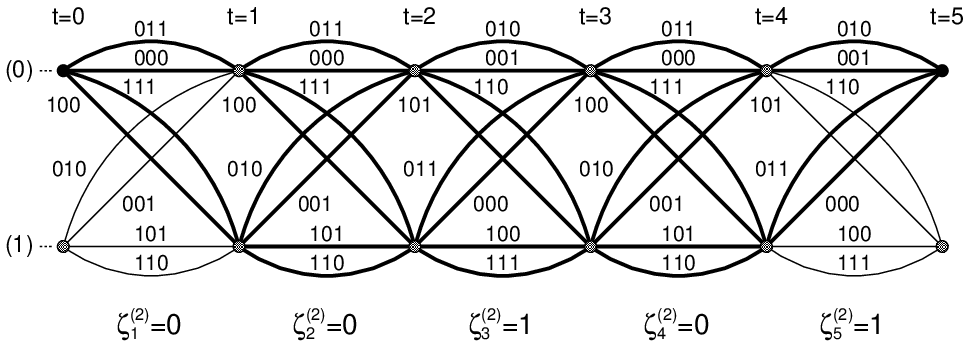}
\end{center}
\caption{A degenerate error trellis based on $H_2^T(D)$.}
\label{Fig.4}
\end{figure}

Now, partition $H^T(D)$ vertically into two submatrices, i.e.,
\begin{eqnarray}
H^T(D) &=& \left(
\begin{array}{cc}
1+D& D \\
D& 1 \\
1+D& 1 
\end{array}
\right) \nonumber \\
&\stackrel{\triangle}{=}& (H_1^T(D), H_2^T(D)) .
\end{eqnarray}
As stated above, the original error paths $\{\mbox{\boldmath $e$}_k\}$ can be equally represented using the error trellis based on $H_2^T(D)$. The set of error-trellis modules associated with $H_2^T(D)$ and the overall error trellis are shown in Fig.3 and Fig.4, respectively. Since $\sigma_5^{(2)}=0$, the error trellis in Fig.4 is terminated in state $(0)$ at time $5$. We remark that the additional error paths not allowed in the original error trellis are included in the error trellis in Fig.4. This is because the condition $\mbox{\boldmath $e$}_kH_1^T(D)=\zeta_k^{(1)}$ is not taken into account for error-trellis construction.

\section{Structure of Degenerate Error-Trellis Modules}
\subsection{Observer Canonical Form of a Syndrome Former and Degenerate Error-Trellis Modules}
\begin{figure}[tb]
\begin{center}
\includegraphics[width=5.0cm,clip]{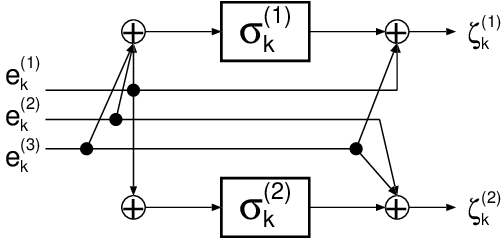}
\end{center}
\caption{Adjoint-obvious realization (observer canonical form) of the syndrome former $H^T(D)$ in (3).}
\label{Fig.5}
\end{figure}

Consider the trellis modules with values $\zeta_k^{(1)}\zeta_k^{(2)}=00$ and $10$ in Fig.1. We see that these trellis modules degenerate into the trellis module with value $\zeta_k=0$ in Fig.3, if two states with the same second component are identified. In the same way, two trellis modules with values $\zeta_k^{(1)}\zeta_k^{(2)}=01$ and $11$ in Fig.1 degenerate into the trellis module with value $\zeta_k=1$ in Fig.3. That is, the trellis modules associated with $H_2^T(D)$ are obtained from the trellis modules associated with $H^T(D)$. In this case, two trellis modules with the same syndrome bit $\zeta_k^{(2)}$ degenerate into the identical trellis module with value $\zeta_k^{(2)}$. This fact is derived from an adjoint-obvious realization (observer canonical form [3]) of the syndrome former $H^T(D)$ (see Fig.5). Note that two pairs $(\sigma_k^{(1)}, \zeta_k^{(1)})$ and $(\sigma_k^{(2)}, \zeta_k^{(2)})$ are independent of each other, from which the above observation is obtained. In general, error-trellis module degeneration is closely related to a realization of the syndrome former. Let
\begin{equation}
H^T(D)=(\mbox{\boldmath $h$}_1^T(D), \mbox{\boldmath $h$}_2^T(D), \cdots, \mbox{\boldmath $h$}_m^T(D))
\end{equation}
and consider the observer canonical form [3] of the syndrome former $H^T(D)$ (cf. Fig.5). Denote by $\nu^{(q)}$ and $L$ the maximum degree among the polynomials of $\mbox{\boldmath $h$}_q(D)~(1\leq q \leq m)$ and the maximum degree among the entries of $H(D)$, respectively. Since equations $\mbox{\boldmath $e$}_k\mbox{\boldmath $h$}_q^T(D)=\zeta_k^{(q)}~(1\leq q \leq m)$ are independent of one another, a realization of $H^T(D)$ is obtained by combining the realizations of $\mbox{\boldmath $h$}_q^T(D)$ in a parallel way (cf. Fig.5). Let $\sigma_{kp}^{(q)}~(1\leq p \leq L)$ be the contents of the memory elements in the realization of $\mbox{\boldmath $h$}_q^T(D)$. (For any fixed $q$, $\sigma_{k1}^{(q)}$ represents the memory element which is closest to the output $\zeta_k^{(q)}$.) Using $\sigma_{kp}^{(q)}$, the state [9] of $H^T(D)$ is defined as
\begin{equation}
\mbox{\boldmath $\sigma$}_k=(\sigma_{k1}^{(1)}, \cdots, \sigma_{k1}^{(m)}, \cdots, \sigma_{kL}^{(1)}, \cdots, \sigma_{kL}^{(m)}) .
\end{equation}
Here, if a memory element is missing, the corresponding $\sigma_{kp}^{(q)}$ is set to zero. Hence, the size of $\mbox{\boldmath $\sigma$}_k$ is $\nu=\nu^{(1)}+\cdots +\nu^{(m)}$. In the following, $H^T(D)$ is assumed to have the form
$$
H^T(D)=(H_1^T(D), H_2^T(D)) ,
$$
where the same conditions as in Section II are assumed. Denote by $\mbox{\boldmath $\sigma$}_k^{(i)}$ and $\mbox{\boldmath $\zeta$}_k^{(i)}$ the state and syndrome at time $k$ corresponding to $H_i^T(D)~(i=1, 2)$, respectively. From the definition of $\mbox{\boldmath $\sigma$}_k$, $\mbox{\boldmath $\sigma$}_k^{(i)}~(i=1, 2)$ may not consist of consecutive components of the state $\mbox{\boldmath $\sigma$}_k$. However, the total of $\mbox{\boldmath $\sigma$}_k^{(i)}~(i=1, 2)$ coincides with $\mbox{\boldmath $\sigma$}_k$ as a whole. We have the following.
\newtheorem{pro}{Proposition}
\begin{pro}
The error-trellis module with value $\mbox{\boldmath $\zeta$}_k^{(1)}$ associated with $H_1^T(D)$ is obtained from the error-trellis module with value $\mbox{\boldmath $\zeta$}_k^{(1)}\mbox{\boldmath $\zeta$}_k^{(2)}$ associated with $H^T(D)$ by reducing the state $(\mbox{\boldmath $\sigma$}_k^{(1)}, \mbox{\boldmath $\sigma$}_k^{(2)})$ to the state $\mbox{\boldmath $\sigma$}_k^{(1)}$. In this case, the trellis modules with the same syndrome component $\mbox{\boldmath $\zeta$}_k^{(1)}$ degenerate into the identical trellis module with value $\mbox{\boldmath $\zeta$}_k^{(1)}$.
\end{pro}
\begin{IEEEproof}
Assume that the syndrome former $H^T(D)$ is in state $\mbox{\boldmath $\sigma$}_{k-1}=(\mbox{\boldmath $\sigma$}_{k-1}^{(1)}, \mbox{\boldmath $\sigma$}_{k-1}^{(2)})$ and an error $\mbox{\boldmath $e$}_k$ is inputted to the syndrome former. Then it goes to state $\mbox{\boldmath $\sigma$}_k=(\mbox{\boldmath $\sigma$}_k^{(1)}, \mbox{\boldmath $\sigma$}_k^{(2)})$ and outputs the syndrome $\mbox{\boldmath $\zeta$}_k=(\mbox{\boldmath $\zeta$}_k^{(1)}, \mbox{\boldmath $\zeta$}_k^{(2)})$. Here, owing to the independence of the first component from the second one, if $\mbox{\boldmath $e$}_k$ is inputted to the syndrome former $H_1^T(D)$, then it goes from state $\mbox{\boldmath $\sigma$}_{k-1}^{(1)}$ to state $\mbox{\boldmath $\sigma$}_k^{(1)}$ and outputs the syndrome $\mbox{\boldmath $\zeta$}_k^{(1)}$ independently of the second component. This fact implies that the degenerate trellis module with value $\mbox{\boldmath $\zeta$}_k^{(1)}$ associated with $H_1^T(D)$ is obtained from the trellis module with value $\mbox{\boldmath $\zeta$}_k^{(1)}\mbox{\boldmath $\zeta$}_k^{(2)}$ associated with $H^T(D)$.
\end{IEEEproof}

\subsection{Relationship Between Degenerate Error-Trellis Modules and the Original Error-Trellis Modules}
Consider any branch, e.g., the branch $\mbox{\boldmath $e$}_k=000$ in the trellis module with value $\zeta_k=0$ in Fig.3. Let $\sigma_{k-1}^{(2)}=(0)$ be the state at time $k-1$ of the branch. If this trellis module is obtained from the trellis module with value $\zeta_k^{(1)}\zeta_k^{(2)}=00$, then $\sigma_{k-1}^{(1)}=0$ is used as the first component of the state. On the other hand, if this trellis module is obtained from the trellis module with value $\zeta_k^{(1)}\zeta_k^{(2)}=10$, then $\sigma_{k-1}^{(1)}=1$ is used as the first component of the state. We observe that all possible patterns of $\sigma_{k-1}^{(1)}$, i.e., $0$ and $1$ appear (a pattern of $\sigma_{k-1}^{(1)}$ depends on the error $\mbox{\boldmath $e$}_k$ in general). These observations can be generalized.
\begin{pro}
Consider any branch $\mbox{\boldmath $e$}_k$ in the degenerate trellis module with value $\mbox{\boldmath $\zeta$}_k^{(1)}$. Denote by $\mbox{\boldmath $\sigma$}_{k-1}^{(1)}$ the state at time $k-1$ of the branch. Assume that this degenerate trellis module is obtained from the trellis module with value $\mbox{\boldmath $\zeta$}_k^{(1)}\mbox{\boldmath $\zeta$}_k^{(2)}$. Then $2^{\nu_2}/2^{m_2}$ patterns of $\mbox{\boldmath $\sigma$}_{k-1}^{(2)}$ are used when viewed from the original trellis module. These sets of patterns are disjoint for different values of $\mbox{\boldmath $\zeta$}_k^{(2)}$ and every possible patten of $\mbox{\boldmath $\sigma$}_{k-1}^{(2)}$ appears when $\mbox{\boldmath $\zeta$}_k^{(2)}$ runs over the $2^{m_2}$ values. The total number of patterns of $\mbox{\boldmath $\sigma$}_{k-1}^{(2)}$ is $(2^{\nu_2}/2^{m_2})\times 2^{m_2}=2^{\nu_2}$.
\end{pro}
\begin{IEEEproof}
Consider any branch $\mbox{\boldmath $e$}_k:\mbox{\boldmath $\sigma$}_{k-1}^{(1)}\rightarrow \mbox{\boldmath $\sigma$}_k^{(1)}$ in the degenerate trellis module with value $\mbox{\boldmath $\zeta$}_k^{(1)}$. When viewed from the original trellis module, given the $\mbox{\boldmath $e$}_k$, the syndrome former goes from $\mbox{\boldmath $\sigma$}_{k-1}=(\mbox{\boldmath $\sigma$}_{k-1}^{(1)}, \mbox{\boldmath $\sigma$}_{k-1}^{(2)})$ to $\mbox{\boldmath $\sigma$}_k=(\mbox{\boldmath $\sigma$}_k^{(1)}, \mbox{\boldmath $\sigma$}_k^{(2)})$ and outputs the syndrome $\mbox{\boldmath $\zeta$}_k=(\mbox{\boldmath $\zeta$}_k^{(1)}, \mbox{\boldmath $\zeta$}_k^{(2)})$. Here, the mapping: $\mbox{\boldmath $\zeta$}_k^{(2)} \mapsto \mbox{\boldmath $\sigma$}_{k-1}^{(2)}$ is one-to-one. Since the number of values of $\mbox{\boldmath $\zeta$}_k^{(2)}$ is $2^{m_2}$, the number of patterns of $\mbox{\boldmath $\sigma$}_{k-1}^{(2)}$ corresponding to a particular value of $\mbox{\boldmath $\zeta$}_k^{(2)}$ is $2^{\nu_2}/2^{m_2}$.
\end{IEEEproof}
\newtheorem{cor}{Corollary}
\begin{cor}
Assume that a degenerate trellis module with value $\mbox{\boldmath $\zeta$}_k^{(1)}$ is obtained from a particular trellis module with value $\mbox{\boldmath $\zeta$}_k^{(1)}\mbox{\boldmath $\zeta$}_k^{(2)}$ and consider the state $\mbox{\boldmath $\sigma$}_k^{(1)}$ at time $k$ in the degenerate trellis module. Then the number of branches entering state $\mbox{\boldmath $\sigma$}_k^{(1)}$ is $2^{n-m}\times 2^{\nu_2}$ when viewed from the corresponding original trellis module.
\end{cor}
\begin{IEEEproof}
A branch connecting state $\mbox{\boldmath $\sigma$}_{k-1}^{(1)}$ to state $\mbox{\boldmath $\sigma$}_k^{(1)}$ in a degenerate trellis module corresponds to $2^{\nu_2}/2^{m_2}$ branches when viewed from the original trellis module. Also, since the size of $H_1(D)$ is $m_1\times n$, the number of branches entering state $\mbox{\boldmath $\sigma$}_k^{(1)}$ is $2^{n-m_1}$.
\end{IEEEproof}
\begin{cor}
Assume that a degenerate trellis module with value $\mbox{\boldmath $\zeta$}_k^{(1)}$ is obtained from a particular trellis module with value $\mbox{\boldmath $\zeta$}_k^{(1)}\mbox{\boldmath $\zeta$}_k^{(2)}$. In this case, the manner of branches entering the state $\mbox{\boldmath $\sigma$}_k^{(1)}$ in the degenerate trellis module is equivalent to the manner of $2^{n-m}$ branches entering each of the $2^{\nu_2}$ extended states $\mbox{\boldmath $\sigma$}_k=(\mbox{\boldmath $\sigma$}_k^{(1)}, \mbox{\boldmath $\sigma$}_k^{(2)})$.
\end{cor}
\begin{IEEEproof}
The number of original states $\mbox{\boldmath $\sigma$}_k$ corresponding to the degenerate state $\mbox{\boldmath $\sigma$}_k^{(1)}$ is $2^{\nu_2}$. Also, since the size of $H(D)$ is $m\times n$, the number of branches entering state $\mbox{\boldmath $\sigma$}_k$ in the original trellis module is $2^{n-m}$.
\end{IEEEproof}

\section{Decoding Based on a Degenerate Error Trellis}
\subsection{Decoding Method}
Consider the decoding based on an error trellis constructed from a submatrix $H_1(D)$ of $H(D)$. In this method, the decoding is carried out by restoring the original trellis modules using Proposition 1. For the purpose, the syndrome-subsequence $\{\mbox{\boldmath $\zeta$}_k^{(2)}\}$ computed from $H_2^T(D)$ is utilized as auxiliary information for decoding. Let $\mbox{\boldmath $\sigma$}_{k-1}^{(1)}$ be the initial state of any branch in the degenerate trellis module. Since there are $2^{\nu_2}$ possibilities with respect to the patterns of $\mbox{\boldmath $\sigma$}_{k-1}^{(2)}$ (cf. Proposition 2), we retain $2^{\nu_2}$ survivors for the degenerate state $\mbox{\boldmath $\sigma$}_{k-1}^{(1)}$. The decoding procedure is given as follows.
\par
ƒ{\bf Decoding based on a degenerate trellis}"
\par
{\it Step 1:} Consider any state $\mbox{\boldmath $\sigma$}_k^{(1)}$ at time $k$ in the degenerate trellis module. The state $\mbox{\boldmath $\sigma$}_k^{(1)}$ has $2^{n-m_1}$ incoming branches. Take one such incoming branch. Denote by $\mbox{\boldmath $\sigma$}_{k-1}^{(1)}$ the state at time $k-1$ of the branch. Note that $2^{\nu_2}$ survivors are retained for the state $\mbox{\boldmath $\sigma$}_{k-1}^{(1)}$. Extend these survivors along the branch under consideration. Then we have $2^{\nu_2}$ extended paths. Repeat this procedure for each branch entering state $\mbox{\boldmath $\sigma$}_k^{(1)}$. As a result, we have $2^{\nu_2}\times 2^{n-m_1}$ extended paths in total.
\par
{\it Step 2:} Discard the extended paths which are not consistent with the value of $\mbox{\boldmath $\zeta$}_k^{(2)}$. The number of remaining paths is
$$
(2^{\nu_2}\times 2^{n-m_1})/2^{m_2}=2^{n-m}\times 2^{\nu_2} .
$$
\par
{\it Step 3:} Classify the remaining paths into $2^{\nu_2}$ groups according to the patterns of $\mbox{\boldmath $\sigma$}_k^{(2)}$, where each group contains $2^{n-m}$ paths. Then the best path among each group is selected and the $2^{\nu_2}$ survivors for the state $\mbox{\boldmath $\sigma$}_k^{(1)}$ are newly determined.
\par
{\it Step 4:} Repeat Step 1 $\sim$ Step 3 for each state $\mbox{\boldmath $\sigma$}_k^{(1)}$.
\par
{\it Remark:} Step 2 and Step 3 correspond to Corollary 1 and Corollary 2 in Section III-B, respectively.

\subsection{Decoding Complexity}
We evaluate the decoding complexity based on the error trellis associated with $H_1^T(D)$. We have the following.
\begin{pro}
The total number of survivors required in the decoding based on $H_1^T(D)$ is $2^{\nu}$.
\end{pro}
\begin{IEEEproof}
$2^{\nu_2}$ survivors are retained for each state in the trellis based on $H_1^T(D)$.
\end{IEEEproof}
\begin{pro}
The number of compare-and-select computations required in the decoding based on $H_1^T(D)$ is $2^{\nu}$, where one path is selected from among $2^{n-m}$ paths in the computation.
\end{pro}
\begin{IEEEproof}
A direct consequence of Step 3 in the decoding procedure.
\end{IEEEproof}
In the proposed method based on $H_1^T(D)$, a kind of {\it list decoding} [4] is required. However, the decoding complexity of the proposed method remains unchanged, except for the additional complexity of discarding the extended paths which are not consistent with the value of $\mbox{\boldmath $\zeta$}_k^{(2)}$.

\section{Decoding Based on a 1-State Error Trellis}
\subsection{Check Matrices Based on Circulant Matrices and 1-State Error Trellises}
\begin{figure}[tb]
\begin{center}
\includegraphics[width=8.8cm,clip]{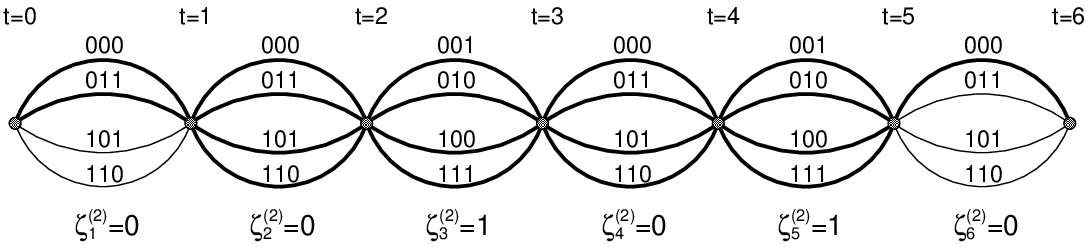}
\end{center}
\caption{A 1-state error trellis based on $H_2^{'T}(D)$.}
\label{Fig.6}
\end{figure}
Tanner et al. [10] proposed LDPC convolutional codes defined by check matrices whose entries are all monomials. As an example [10, {\it Example 7}], take
\begin{equation}
H(D)=\left(
\begin{array}{ccc}
1& D& D^3 \\
D^3& D^2& 1 
\end{array}
\right) ,
\end{equation}
where a common factor in each row has been removed. ({\it Remark:} The above $H(D)$ is not {\it basic} and then not canonical. However, the preceding argument is also effective.) In this case, if we choose any row of $H(D)$ for the purpose of constructing an error trellis for decoding, then a 1-state error trellis is obtained by factoring out a factor $D^l$ from each entry of the row. In order to clarify the idea, again consider the check matrix $H(D)$ given in (3). Since the first entry of $H_2(D)=(D, 1, 1)$ has a factor $D$, we can apply the error-trellis construction method in [9]. Let $e_k^{'(1)}\stackrel{\triangle}{=}De_k^{(1)}=e_{k-1}^{(1)}$ and define as $\mbox{\boldmath $e$}_k'\stackrel{\triangle}{=}(e_k^{'(1)}, e_k^{(2)}, e_k^{(3)})$. We have
\begin{eqnarray}
\mbox{\boldmath $\zeta$}_k &=& (e_k^{'(1)}, e_k^{(2)}, e_k^{(3)})\left(
\begin{array}{cc}
1+D^{-1}& 1 \\
D& 1 \\
1+D& 1 
\end{array}
\right) \nonumber \\
&\stackrel{\triangle}{=}& \mbox{\boldmath $e$}_k'(H_1^{'T}(D), H_2^{'T}(D))=\mbox{\boldmath $e$}_k'H^{'T}(D) .
\end{eqnarray}
Since $H_2'(D)=(1, 1, 1)$, the error paths associated with $H_2^T(D)$ are represented using a 1-state error trellis [9]. A 1-state error trellis equivalent to the one in Fig.4 is shown in Fig.6. Since the two trellises in Fig.4 and Fig.6 are equivalent, the decoding method stated in Section IV-A can be mapped on the error trellis in Fig.6. Actually, we can regard $H'(D)$ as a given check matrix and then apply our method to it.  Since the constraint length of $H_1'(D)=(1+D^{-1}, D, 1+D)$ is $2$, if $4$ survivors are retained, then ML decoding is accomplished. Applying the method, we have the decoded path $\mbox{\boldmath $\hat e$}'=000~000~100~000~100~000$ (cf. $\mbox{\boldmath $\hat e$}$). Though the ML decoding is realized using the above method, the decoding complexity is not reduced compared to the original one. In the following, therefore, we propose a simplified decoding method taking into account a feature of error trellises.

\subsection{Likelihood Distribution of the Trellis-State and Sub-Optimal Decoding}
Note that only eight paths in the error trellis in Fig.6 are admissible. As a result, the {\it principle of optimality} does not hold for the error trellis in Fig.6. That is, the error path with minimum weight which is consistent with the values of $\{\zeta_t^{(1)}\}_{t=1}^k$ is not necessarily consistent with the values of $\{\zeta_t^{(1)}\}_{t=1}^N$ at the final time $N$. However, it is reasonable to imagine that an error path with low weight which is consistent with $\{\zeta_t^{(1)}\}_{t=1}^k$ is likely to become the overall optimal path at time $N$. Here, take notice of a feature of error trellises. Again consider the check matrix $H(D)$ given in (7). Let $e_k^{'(2)}\stackrel{\triangle}{=}De_k^{(2)}=e_{k-1}^{(2)}$ and $e_k^{'(3)}\stackrel{\triangle}{=}D^3e_k^{(3)}=e_{k-3}^{(3)}$. Also, define as $\mbox{\boldmath $e$}_k'\stackrel{\triangle}{=}(e_k^{(1)}, e_k^{'(2)}, e_k^{'(3)})$. Then we have
\begin{eqnarray}
\mbox{\boldmath $\zeta$}_k &=& (e_k^{(1)}, e_k^{'(2)}, e_k^{'(3)})\left(
\begin{array}{cl}
1& D^3 \\
1& D \\
1& D^{-3} 
\end{array}
\right) \nonumber \\
&\stackrel{\triangle}{=}& \mbox{\boldmath $e$}_k'(H_1^{'T}(D), H_2^{'T}(D))=\mbox{\boldmath $e$}_k'H^{'T}(D) .
\end{eqnarray}
The time-$k$ state of the error trellis based on $H_2^{'T}(D)$ is expressed as
\begin{eqnarray}
\mbox{\boldmath $\sigma$}_k' &=& (e_{k-2}^{(1)}+e_k^{'(2)},~e_{k-1}^{(1)}+e_{k+1}^{'(2)},~e_k^{(1)}+e_{k+2}^{'(2)}, \nonumber \\
&& e_{k+1}^{(1)}+e_{k+3}^{'(2)},~e_{k+2}^{(1)},~e_{k+3}^{(1)}) .
\end{eqnarray}
Let $\epsilon\stackrel{\triangle}{=}P(e_k^{(i)}=1)$ be the channel crossover probability. Denote by $q_{\bar 0}$ and $q_{\bar i}~(1\leq i \leq 6)$ the probabilities of the all-zero state and the state with $0$'s except for the $i$th entry, respectively. $q_{\bar 0}\sim q_{\bar 6}$ are given as follows.
\begin{eqnarray}
q_{\bar 0} &=& (1-2\epsilon+2\epsilon^2)^4(1-\epsilon)^2 \nonumber \\
q_{\bar 1}\sim q_{\bar 4} &=& 2\epsilon(1-2\epsilon+2\epsilon^2)^3(1-\epsilon)^3 \nonumber \\
q_{\bar 5}\sim q_{\bar 6} &=& \epsilon(1-2\epsilon+2\epsilon^2)^4(1-\epsilon) . \nonumber
\end{eqnarray}
\begin{table}[tb]
\caption{$q_{\bar 0}\sim q_{\bar 6}$ and their sum versus $\epsilon$.}
\label{table:2}
\begin{center}
\begin{tabular}{l*{4}{|l}}
$\epsilon$ & $q_{\bar 0}$ & $q_{\bar 1}\sim q_{\bar 4}$ & $q_{\bar 5}\sim q_{\bar 6}$ & $\sum_{i=0}^6q_{\bar i}$ \\
\hline
$0.1$ & $0.49$ & $0.02$ & $0.01$ & $0.59$ \\
$0.05$ & $0.6225$ & $0.055$ & $0.0275$ & $0.8975$ \\
$0.01$ & $0.9049$ & $0.0182$ & $0.0091$ & $0.9959$ \\
$0.005$ & $0.951225$ & $0.00955$ & $0.004775$ & $0.998975$ \\
$0.001$ & $0.990049$ & $0.001982$ & $0.000991$ & $0.999959$
\end{tabular}
\end{center}
\end{table}

The values of $q_{\bar 0}\sim q_{\bar 6}$ and their sum versus $\epsilon$ are shown in TABLE I (evaluated up to order $\epsilon^2$). In this example, the sum is greater than $0.99$ for $\epsilon \leq 0.01$. We see a likelihood-concentration [7] occurs in the all-zero state and the states with many $0$'s. Hence, we choose as survivors the error paths with $M~(\leq 2^{\nu})$ lowest weights from among the paths which have passed the syndrome test (i.e., M-algorithm [2], [11]). The decoding procedure is given as follows.
\par
ƒ{\bf Decoding based on a 1-state error trellis}"
\par
{\it Step 1:} (Let $H_1(D)=\mbox{\boldmath $h$}_1(D)$.) Suppose that $M$ survivors are retained at time $k+\nu^{(1)}-1$. Extend these survivors by the unit time based on the error trellis associated with $\mbox{\boldmath $h$}_1^{'T}(D)$. We have $M \times 2^{n-1}$ extended paths in total.
\par
{\it Step 2:} Discard the extended paths which are not consistent with the value of $\mbox{\boldmath $\zeta$}_k^{(2)}$. The number of remaining paths is $M \times 2^{n-1}/2^{m-1}=2^{n-m}\times M$.
\par
{\it Step 3:} Order the remaining paths by their metrics (weights) and select the best $M$ paths as the survivors at time $k+\nu^{(1)}$.
\par
A Viterbi algorithm making use of a likelihood distribution of the state in a code trellis was proposed in [7]. A similar idea combined with the M-algorithm is described in [11]. It is shown [11] that $M$ can be reduced to $16$ within a very small degradation compared to ML decoding for a $(2, 1, 8)$ convolutional code (cf. the number of trellis states $S=256$). (It is stated [2] that $M\approx \sqrt{S}$ is asymptotically optimal for large codes.) Since a likelihood-concentration occurs in error trellises, the method is also effective for error trellises.

\section{Conclusion}
We have presented a decoding method using an error trellis constructed based on a submatrix of a given check matrix. In this method, the given check matrix is partitioned into two submatrices and one is used for error-trellis construction, whereas the other is used for generation of auxiliary information for decoding. We have shown that the ML error path is correctly decoded using the degenerate error trellis based on the former submatrix, if the syndrome-subsequence computed from the latter is utilized as side information for decoding. Next, we have applied the method to check matrices with monomial entries proposed by Tanner et al. It is shown that by choosing any row of the check matrix as the submatrix for error-trellis construction, a $1$-state error trellis is obtained. Taking into account a significant feature (i.e., a likelihood-concentration on the all-zero state and the states with many $0$'s), we have proposed a simplified decoding method (M-algorithm) based on a $1$-state error trellis, from which decoding-complexity reduction is realized.






%

\end{document}